\documentclass[prb,preprint]{revtex4}

\usepackage{graphicx}

\usepackage{amsmath}
\usepackage{comment}

\newcommand{\D}{\Delta}
\newcommand{\dos}{density of states}
\newcommand{\md}{molecular dynamics}
\newcommand{\mc}{Monte Carlo}
\newcommand{\lb}{{<}}
\newcommand{\rb}{{>}}
\newcommand{\gee}{g(E)}

\newcommand{\hee}{H(E)}
\newcommand{\WL}{Wang-Landau}
\newcommand{\geet}{\tilde g(E)}
\newcommand{\gt}{{\tilde g}}

\newcommand{\tran}{\pi}
\newcommand{\de}{\Delta E}

\begin{document}

\title{Teaching statistical physics by thinking about models and algorithms}
\author{Jan Tobochnik}
\email{jant@kzoo.edu}
\affiliation{Department of Physics, Kalamazoo College, Kalamazoo, Michigan 49006}
\author{Harvey Gould}
\email{hgould@clarku.edu}
\affiliation{Department of Physics, Clark University, Worcester,
Massachusetts 01610}

%\date{\today}

\begin{abstract}
We discuss several ways of illustrating fundamental concepts in statistical and thermal physics by considering various models and algorithms. We emphasize the importance of replacing students' incomplete mental images by models that are physically accurate. In some cases it is sufficient to discuss the results of an algorithm or the behavior of a model rather than having students write a program.
\end{abstract}

\maketitle

\section{Introduction}

Mathematics is both the language of physics and a calculational tool. For example, the statements ${\bf a} = {\bf F}/m$ and $\boldmath{\nabla}\cdot{\bf B} =0$ express the ideas that acceleration is the result of forces and magnetic field lines exist only as closed loops. The fundamental thermodynamic relation $dS = (1/T)dU + (P/T)dV - (\mu/T)dN$ implies that there exists an entropy function that depends on the internal energy $U$, the volume $V$, and the number of particles $N$. It also tells us that the temperature $T$, the pressure $P$, and the chemical potential $\mu$ determine how the entropy changes. 

Although textbooks and lectures describe the meaning of mathematical relations in physics, students frequently do not understand their meaning because there are few related activities that undergraduate students can do. Instead students frequently use mathematics as a calculational tool for problems which lead to little physical understanding. To introductory students ${\bf a} = {\bf F}/m$ is just one of many algebraic relations which needs to be manipulated. For more advanced students it is a differential equation which needs to be solved.

The availability of symbolic manipulation software and calculators that can do much of the mathematical manipulations which students have been traditionally asked to do means that instructors need to think carefully about what are the appropriate activities for students. How many of the problems at the back of textbook chapters teach useful skills or help students learn physics? Which skills are important? How much, if any, understanding is lost if more of the calculations are done with the aid of a computer? Is most of the physics in the setup of the problem and in the analysis of the results? Should we spend more time teaching students to use software packages more effectively? 

In addition to challenging the way we teach the traditional physics curriculum, there is a growing interest in including more computation into the curriculum.
Many physics teachers view computation as another tool analogous to mathematical tools such as those used to solve algebraic and differential equations. The theory is the same and computation is needed only when exact solutions are not available. Because the latter provide useful illustrations of the theory, computation need be only a minor part of the undergraduate (and even graduate) curriculum. This situation is the common practice at most institutions. 

A popular rationale for more computation in the curriculum is that it allows us to do more realistic problems, is important in scientific research, and provides a useful tool for later employment. These reasons are all valid, but they might be less important than they seem. Although computation can allow us to consider more realistic problems, the consideration of such problems usually requires a much greater understanding of the system of interest than most undergraduates have the background and time to learn. Knowing a programming language is useful in employment, but might become less important as more and more work is done by higher level languages targeted toward specific applications. And even though computation is ubiquitous in research, more of it is being done using software packages. Just as few experimentalists need to know the details of how electronic instrumentation is constructed, few scientists need to know the details of how software packages are written. Moreover there usually is not enough time for students to write their own programs except in specialized courses in computational physics and simulation.

In this paper we argue that computation should be incorporated into the curriculum because it can elucidate the physics. As for mathematics, computation is both a language and a tool. In analogy to models expressed in mathematical statements there are models expressed as algorithms. In many cases the algorithms are explicit implementations of the mathematics. For example, writing a differential equation is almost the same as writing a sequence of rate equations for the variables in a computer program. An advantage of the computational approach is that it is necessary to be explicit about which symbols represent variables and which represent initial conditions and parameters.

In other cases the algorithm does not look like a traditional mathematical statement. For example, the Monte Carlo algorithms used in statistical physics do not look like expressions for the partition function or the free energy. There also are models such as cellular automata that have no counterpart in traditional mathematics.

In the following we will discuss some examples of how the consideration of algorithms and models can help students and instructors understand some fundamental concepts in physics. Because computation has had a great influence on statistical physics, we will focus our attention on this area. It is also the area in which we have the most expertise.

\section{\label{approach}Approach to equilibrium}

A basic understanding of probability is necessary for understanding the statistical behavior underlying thermal systems. Consider the following question, which can be stated simply in terms of an algorithm.\cite{eisberg} Imagine two bags and a large number of balls. All the balls are initially placed in bag one. Choose a ball at random and move it to the other bag. After many moves, how many balls are in each bag? Many students will say that the first bag will have more balls on the average.\cite{interviews} In this case it might be useful to show a simulation of the ball swapping process.\cite{stp} It is then a good idea to ask students to sketch the number of balls in one bag as a function of time before showing the simulation. We can then discuss why the mean number of balls in each bag eventually becomes time independent. In particular, we can illustrate the presence of fluctuations during relaxation and in equilibrium and how the results depend on the number of particles. We can illustrate many of the basic features of thermal systems including the concepts of microstate and macrostate, the history independence of equilibrium (the equilibrium macrostate is independent of the initial conditions), ensemble and time averages, and the need for a statistical approach. We can also discuss why fluctuations in macroscopic systems will be negligible for most thermal quantities. 

The approach to equilibrium can be repeated with a molecular dynamics simulation (see Sec.~\ref{sec:md}) in which particles are initially confined to one part of a container. After the release of an internal constraint the particles eventually are uniformly distributed on the average throughout the container. It is important in this example to show the students the basic algorithm and some simple computer code, so that they are convinced that there is no explicit ``force" pushing the system toward equilibrium, but rather that equilibrium is a result of random events. 
The idea is to make explicit that the general behavior of systems with many particles is independent of the details of the inter-particle interactions. 

\section{\label{sec:md}Molecular Dynamics}

If students are asked what happens to the temperature when a gas is compressed, they will likely say that it increases. Their microscopic explanation will likely be that the molecules rub against each other and ``give off heat.''\cite{heron,interviews} These students have a mental model which is similar to a system of marbles with inelastic collisions. 

This granular matter model of a gas and liquid is appropriate for the types of non-thermal macroscopic systems that students experience in everyday life such as cereals, rice, and sand. For molecular systems the collisions are elastic, and a hard sphere model\cite{aw} with elastic collisions is useful for understanding the properties of fluids and solids. Thus, student intuition has some value, but is limited in its ability to account for much of the phenomena of thermal systems. In particular, the only relevance of the temperature in hard sphere models is to set the the natural energy scale. 

A concrete and realistic model of thermal systems is provided by thinking about molecular dynamics with continuous inter-particle potentials for a system with a fixed number of particles and fixed volume.\cite{md} Each particle is subject to the force of every other particle in the system, with nearby particles usually providing most of the force. The dynamics of the $i$th particle of mass $m_i$ is determined by Newton's second law, ${\bf a}_i = {\bf F}_{{\rm net},i}/m_i$, which can be integrated numerically to obtain the position and velocity of each particle.\cite{noteuler} Students should be asked to think about the appropriate inter-molecular force law and what would happen in the simulation and a real thermal system. For example, what is different about a gas and a liquid? This question will lead students to conclude that there must be an attractive contribution to the force law. Also systems do not completely collapse and thus there must be a repulsive part as well. Students can then be led to conclude that the force law must look something like the force law derivable from the Lennard-Jones potential. 

Next ask students about pressure, and you will likely not receive a coherent answer. Even if students know that pressure is force per unit area, it is likely that they will not be able to explain what that means in terms of a microscopic model of a gas. The simplest way to determine the pressure is to compute the momentum transfer across a surface. Without doing the simulation we can see that because momentum is proportional to velocity, faster particles will lead to greater momentum transfer per unit time. Then we can discuss what other macroscopic quantities increase with particle speed. Students will typically bring up temperature because they associate temperature with kinetic energy. From this discussion they can conclude that the pressure and temperature depend on the mean speed of the particles. Also, if the density increases at a given temperature, more particles will cross a given surface, and hence the pressure increases with the density. Students can then conclude that there are two independent quantities, the density and the temperature, which determine the pressure.

Because the total energy is conserved in an isolated container, students can understand that the total energy consists of both potential and kinetic energy. At this point some discussion is needed to help students understand that there is potential energy that has nothing to do with gravity. Once that understanding is reached, it is easy to understand that the kinetic energy does not remain constant, and thus there will be small fluctuations in the temperature. Similar reasoning can lead to the idea of pressure fluctuations. Thus, by imagining the particles in a molecular dynamics simulation the concept of thermal fluctuations can be inferred. This result can be related to the fluctuations discussed in the ball swapping model in Sec.~\ref{approach}. 

What is the role of temperature? We will discuss temperature again in Sec.~\ref{sec:demon}, but here we consider how molecular dynamics can be used to think about its meaning. Imagine two solids such that initially the particles in one solid are moving much faster than the particles in the other solid. The two solids are placed in thermal contact so that the particles interact with each other across the boundary between the two solids. This model provides a concrete realization of thermal contact. Particles at the boundary will exchange momentum and energy with each other. The faster particles will give energy to the slower ones and energy will be transferred from the solid with the faster particles to the one with the slower particles. The net energy transfer will cease when all the particles have the same mean kinetic energy per particle, but not the same total energy per particle, potential energy per particle or mean speed. Thus by thinking about this system we can gain insight about the connection between kinetic energy and temperature. More importantly, we can emphasize that temperature is the quantity that becomes the same when two systems are placed in thermal contact.

Our experience has been that a consideration of molecular dynamics in various contexts over several weeks leads to most students replacing their granular matter model by one in which energy is conserved.

\section{\label{sec:demon}An ideal thermometer}

In a previous paper\cite{demon} we discussed the {\it demon algorithm}, which can be used as a model of an ideal thermometer and as a chemical potential meter. In a computer simulation the demon is an extra degree of freedom which can exchange energy or particles with a system. Energy is exchanged by making a random change in one or more degrees of freedom of the system. If the change increases the energy, the change is accepted only if the demon has enough energy to give the system. If the trial move decreases the energy of the system, the extra energy is given to the demon. In a similar way the demon can exchange particles with the system. The energy distribution of the demon for a given number of particles in the system is the Boltzmann distribution from which the temperature can be extracted because the demon and system are in thermal equilibrium. If particle and energy exchanges are allowed, the distribution of energy and particles held by the demon is a Gibbs distribution from which the chemical potential can be extracted. 

A discussion of the demon is a good way of introducing the concepts of system, heat bath, and ensembles. The demon algorithm simulates the microcanonical ensemble because the total energy of the system plus the demon is fixed, and the demon is a negligible perturbation of the system. An alternative interpretation is that the demon is the system of interest and the remaining particles play the role of the heat bath as in a canonical ensemble. The demon is unusual because its energy is both the energy of the system and the energy of a microstate. Similarly, the state of the demon is both a microstate and a macrostate. In realistic thermal systems there are many microstates corresponding to a given macrostate, and thus there are distinctions between a macrostate, a microstate, and a single particle state. The demon algorithm provides a concrete example for discussing such concepts. 

\section{Markov chains and the Metropolis algorithm}

One of the most common algorithms used to simulate thermal systems is the Metropolis algorithm.\cite{mc} In this section we discuss the underlying theory behind this algorithm, which will allow us to introduce concepts such as probability distributions, the Boltzmann distribution, Markov chains, the partition function, sampling, and detailed balance. In Sections VI-VIII we will extend these ideas to newer algorithms that are currently of much interest in research.

Most students are familiar with probabilities in the context of dice and other simple games of chance, but they have difficulty understanding what probabilities mean in the context of thermal systems. 

Students are familiar with systems that evolve in time according to deterministic equations such as Newton's second law. We can also consider stochastic processes such as those used in Monte Carlo simulations (see Sec.~\ref{approach}). In this case we consider an ensemble of identical copies of a system. These copies have the same macroscopic parameters, interactions, and constraints, but the microscopic states (such as the positions and momenta of the particles) are different in general. We imagine a probabilistic process which changes the microscopic states of the members of the ensemble. An example is a Markov process for which the probability distribution of the states of the ensemble at time $t$ depends only on the probability distribution of the ensemble at the previous time; that is, the system has no long term memory. A Markov process can be represented by the equation:
\begin{equation}
P(t+\Delta t)= \tran P(t),
\label{markov}
\end{equation}
where $P(t)$ is a column matrix of $n$ entries, each one representing the probability of a microstate; $\tran$ is a matrix of transition probabilities. The $i$th entry of $P(t)$ gives the fraction of the members of the ensemble that are in the $i$th state at time $t$. By repeatedly acting on $P$ with the transition matrix we arrive at the distribution at any later time $t$. If a system is in a stationary state, then 
\begin{equation}
P(t+\Delta t) = P(t). 
\label{stationary}
\end{equation}

How can we be sure that the Markov process we use in our simulations will lead to the theoretically correct equilibrium distribution? We begin with the detailed balance condition
\begin{equation}
\tran_{i \to j}\,P_i = \tran_{j \to i}\,P_j,
\label{generaldetailedbalance}
\end{equation}
where $\tran_{i \to j}$ is the probability of the system going from state $i$ to state $j$, and $P_i$ is the probability that the system is in state $i$. Detailed balance puts a constraint on the transition probabilities. As we will show, if the stochastic process satisfies detailed balance, then a system which is in a stationary state will remain in that state. 

We now show that this property is satisfied for an equilibrium system at temperature $T=1/k\beta$ for which the probability of a microstate is given by the Boltzmann distribution. 
If we substitute the Boltzmann distribution $P_i = e^{-\beta E_i}/Z$ into Eq.~\eqref{generaldetailedbalance}, we obtain
\begin{equation}
\tran_{i \to j}\,e^{-\beta E_i} = \tran_{j \to i}\,e^{-\beta E_j},
\label{detailedbalance}
\end{equation}
where $E_i$ is the energy of state $i$ and $k$ is Boltzmann's constant. Note that the partition function $Z=\sum_i e^{-\beta E_i}$ does not appear in Eq.~\eqref{detailedbalance}. This absence is important because $Z$ is usually not known.

We use Eq.~(\ref{detailedbalance}) and the Boltzmann distribution for the desired stationary state to calculate the right-hand side of each row of Eq.~(\ref{markov}):
\begin{equation}
P_i(t+\Delta t)=\sum_j \tran_{j \to i} P_j(t) .
\label{eachrow}
\end{equation}
We next use the detailed balance condition in Eq.~(\ref{detailedbalance}) to obtain
\begin{equation}
P_i(t+\Delta t) = \sum_j \big[\tran_{i \to j} e^{-\beta(E_i - E_j)}\big]P_j(t).
\label{eachrow2}
\end{equation}
We want to see if an ensemble that is described by the Boltzmann distribution will remain in this distribution. We replace $P_j$ in Eq.~(\ref{eachrow2}) by the Boltzmann distribution $e^{-\beta E_j}/Z$ so that
\begin{equation}
P_i(t+\Delta t) = \sum_j \big[\tran_{i \to j} e^{-\beta(E_i - E_j)}\big]\frac{e^{-\beta E_j}}{Z}.
\label{eachrow3}
\end{equation}
We can simplify Eq.~\eqref{eachrow3} by writing 
\begin{equation}
P_i(t+\Delta t) = \frac{e^{-\beta E_i}} {Z} \sum_j \tran_{i \to j}, 
\label{eachrow4}
\end{equation}
where we have taken the factor that does not depend on $j$ out of the sum. Because the system has probability unity of going from the state $i$ to all possible states $j$, the sum over $j$ must equal unity, and thus Eq.~(\ref{eachrow4}) becomes
\begin{equation}
P_i(t+\Delta t)= \frac{e^{-\beta E_i}}{Z}.
\label{eachrow}
\end{equation}
Thus, we have shown that if we start with an ensemble distributed according to the Boltzmann distribution and use a transition probability that satisfies detailed balance, then the resulting ensemble remains in the Boltzmann distribution. Note that detailed balance is sufficient, but we have not shown that detailed balance is necessary for an ensemble to remain in a stationary state, and it turns out that it is not.

Although detailed balance specifies the ratio $\tran_{i \to j}/\tran_{j \to i}$, it does not specify $\tran_{i \to j}$ itself. There is much freedom in choosing $\tran_{i \to j}$, and the optimum choice depends on the nature of the system being simulated. The earliest and still a popular choice is $\tran_{i \to j} = A_{i\to j}W_{ij}$, where $W_{ij}$ is the probability of making an arbitrary trial move and $A_{i\to j}$ is the Metropolis acceptance probability given by 
\begin{equation} 
A_{i \to j} = \min \Big\{1, e^{-\beta(E_j - E_i)} \Big\}.
\label{metropolis}
\end{equation}
Equation~\eqref{metropolis} can be shown to satisfy detailed balance if $W_{ij} = W_{ji}$ (see Problem~\ref{prob:metro}). For example, for an Ising model with $N$ spins choosing a spin at random implies $W_{ij} = W_{ji} = 1/N$.

Our discussion of probability has been a mixture of some abstract ideas from probability theory and the example of the Metropolis algorithm. The discussion can be made more concrete by considering a specific model, such as the Ising model of magnetism or a Lennard-Jones system of particles.\cite{gtcilj} 

\section{Direct estimation of the Density of States}

The density of states $g(E)$ is defined\cite{densityofstates} so that $g(E)\D E$ is the number of microstates with energy between $E$ and $E+\D E$. For most students the density of states is an abstract quantity and many confuse the density of states of a many body system with the single particle density of states. The following discussion provides an algorithm which makes the meaning of $g(E)$ more concrete. 

If the density of states $g(E)$ is known, we can calculate the mean
energy and other thermodynamic quantities at any temperature from
the relation
\begin{equation}
\lb E \rb = \sum_E E\,P(E),
\label{aveE}
\end{equation}
where the probability $P(E)$ that a system in equilibrium with a heat bath at
temperature
$T$ has energy
$E$ is given by
\begin{equation}
\label{eq:prob}
P(E) = \frac{g(E)\, e^{-\beta E}}{Z},
\end{equation}
and $Z=\sum_E g(E)\, e^{-\beta E}$. 
Hence, the density of states is of much interest.

Extracting the \dos\ from a \md\ simulation or from the Metropolis and similar algorithms is very difficult. For example, suppose that we try to determine $g(E)$ by doing a random walk in
energy space by flipping the spins in the Ising model at random and accepting all
microstates that are obtained. The histogram of the energy, $H(E)$,
the number of visits to each possible energy $E$ of
the system, would converge to $g(E)$ if the walk visited all possible
configurations. In practice, it would be impossible to realize such
a long random walk given the extremely large number of possible configurations in even a small system.

Recently a \mc\ algorithm due to Wang and Landau\cite{wanglandau} has been developed for determining $g(E)$ directly. The idea is to simulate a system by making changes at random, but to sample energies that are more probable less often so that $H(E)$ becomes approximately independent of $E$. The acceptance criteria is given by 
\begin{equation} 
A_{i\to j} = \min \Big\{1, \frac{\tilde g(E_i)}{\tilde g(E_j)} \Big\}.
\label{flat}
\end{equation}
where $\geet$ is the running estimate of $\gee$. This acceptance criteria favors energies which have a low density of states, and because these states are visited more often, the result is that a histogram of the energy of the visited configurations will be approximately independent of $E$ or ``flat.'' It is possible to show (see Problem~\ref{prob:flat}) that the probability of the $i$th microstate is proportional to $1/g(E_i)$ in the stationary state. 

How do we implement Eq.~(\ref{flat}) when we don't know $\gee$, which is the goal of the simulation? The second part of the algorithm is to make an initial guess for $\geet$ and then improve the estimate of $\geet$ as the simulation proceeds. The simplest guess is to let $\geet = 1$ for all $E$ and then to update $\geet$ and $\hee$ after each trial move:
\begin{subequations}
\label{gcalc}
\begin{align}
\gt_{t+1}(E) & = f \gt_t(E), \label{gupdate}\\
\noalign{\noindent or}
\ln{\gt_{t+1}(E)} & = \ln{\gt_t(E)} + \ln{f},
\end{align}
\end{subequations}
with
\begin{equation}
\label{he}
H_{t+1}(E) = H_t(E) + 1,
\end{equation}
and $f> 1$; $t$ represents the number of updates. (The value of $E$ in Eqs.~\eqref{gcalc} and \eqref{he} is unchanged if the trial move is not accepted.) Because $\geet$ increases rapidly with $E$, we need to use $\ln \geet$ to implement this algorithm on a computer.

The combination of Eqs.~\eqref{flat} and \eqref{gcalc} forces the system to visit less explored energy regions due to
the bias in the acceptance probability in Eq.~\eqref{flat}. For example, if the current estimate $\geet$ is too low, moves to states with a lower value of $\geet$ have a greater probability. In this way the values of $\geet$ will gradually approach the ``true'' $\gee$.

Initially we choose $f$ to be large (typically $f=e \approx 2.7$) so that $g(E)$ will change rapidly. After many moves the histogram $H(E)$ will become approximately flat, and we then decrease $f$ so that $\geet$ doesn't change as much. The usual procedure is to let $f \to \sqrt{f}$ and continue the updates of $\geet$ and the changes in $f$ until $f -1 \sim$ O($10^{-8}$).

\begin{figure}[h!]%fig1
\includegraphics[scale=0.6]{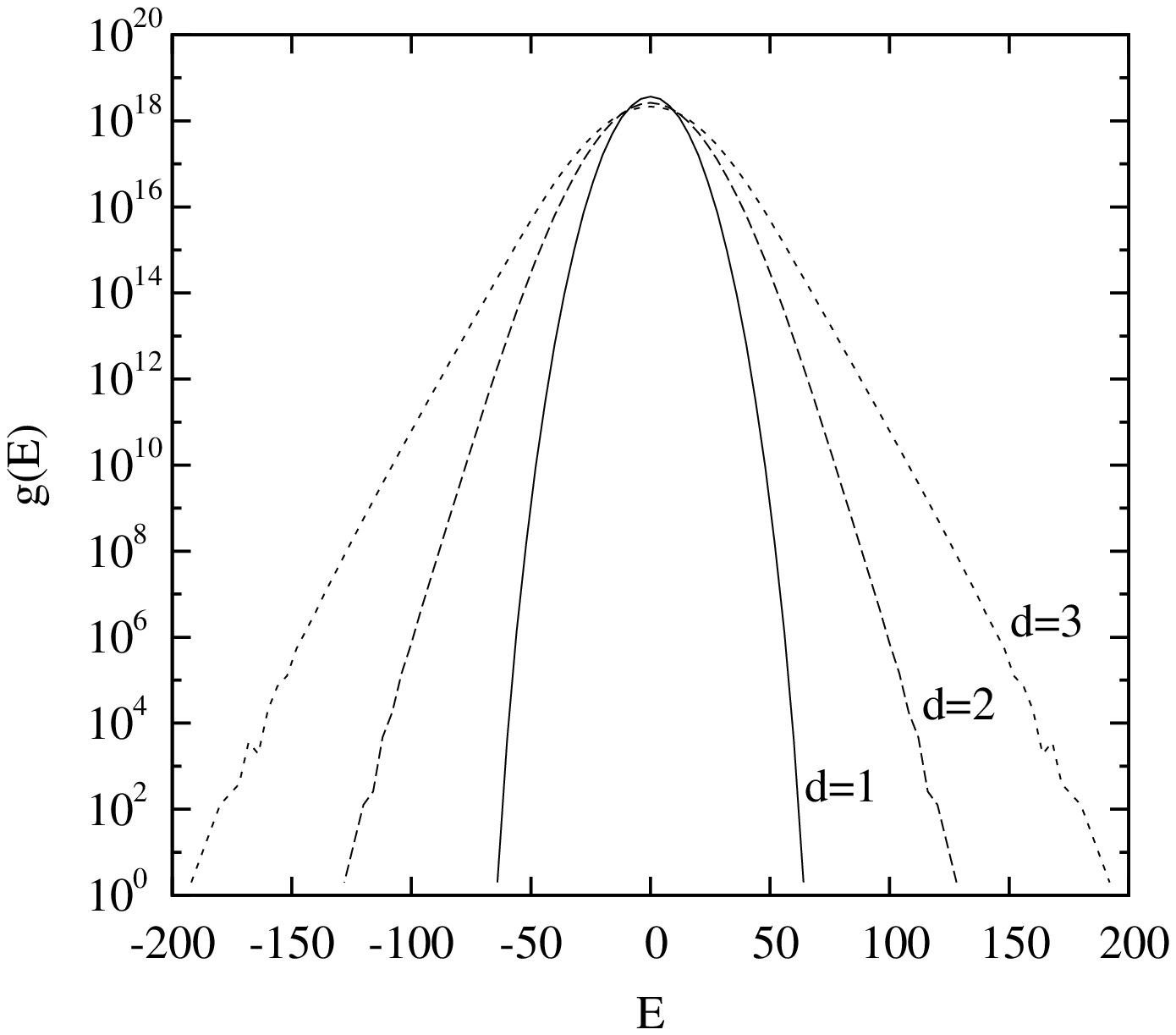}
\vspace{-0.15cm}
\caption{\label{ising123} Semi-log plot of the exact densities of states for $N=64$ spins for one, two, and three dimensions. The results for three dimensions were generated using the method discussed in Ref.~\onlinecite{ising3d}. The results of the Wang-Landau algorithm are indistinguishable from the exact results for these small systems. Note the tiny deviations from a smooth curve at $|E| \approx 120$ for $d=2$ and $|E| \approx 170$ for $d=3$. These are signatures of a phase transition in the thermodynamic limit. For $d=1$ the Ising model does not have a phase transition.}
\end{figure}

\begin{figure}[h!]%fig2
\includegraphics[scale=0.62]{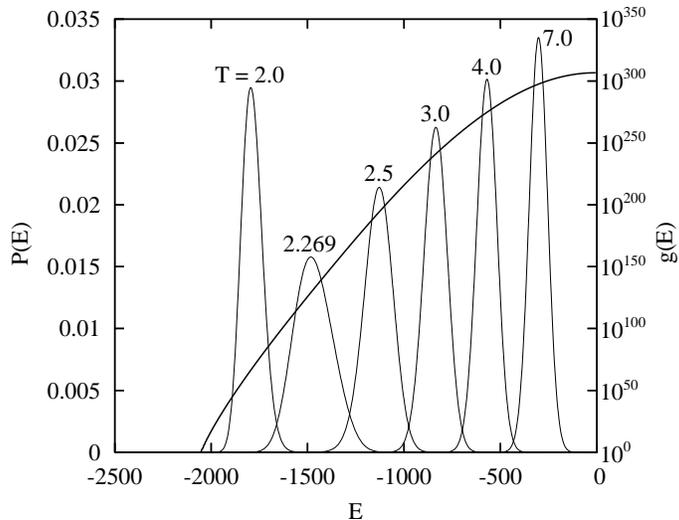}
\vspace{-0.25cm}
\caption{\label{canonical} The probability $P(E)$ of the energy $E$ for the Ising model on a $32 \times 32$ square lattice for various temperatures. Superimposed on the same plot is the density of states.}
\end{figure}

As can be seen in Eq.~(\ref{aveE}) the only quantity that differentiates the mean energy and the specific heat of one system from another is the density of states. However, the density of states of many systems is qualitatively similar. This similarity is illustrated in Fig.~\ref{ising123}, which shows the density of states for 64 Ising spins in one, two, and three dimensions. More insight can be gained from the plot in Fig.~\ref{canonical} of the probability $P(E)$ superimposed on a plot of $\ln \gee$. The plot shows the range of values of $E$ that are important for a given temperature. The competition between the increase of $g(E)$ with $E$ and the decrease of the Boltzmann factor leads to the peak in $P(E)$. 

One of the interesting features of the Wang-Landau algorithm is that it samples states that are of little interest for thermal systems, unlike the Metropolis algorithm which rarely samples such states. For example, an application of the \WL\ algorithm to the Ising model would lead to a parabolic-like curve for $\ln{g(E)}$ with a maximum at $E= 0$ as shown in Fig.~\ref{ising123}. Positive energy states, with most of the nearest neighbor spins of opposite sign, are present. Because large positive energy states would not be observed in a Metropolis simulation of the Ising model, it is clear that we are measuring a temperature independent property that depends only on the nature of the system. 

\section{Direct estimation of T(E) in the microcanonical ensemble\label{kim}}

Another way to determine $\gee$ is to exploit the correspondence between the density of states
and the thermodynamic temperature $T(E)$
and to update the latter rather than $\gee$.\cite{stmc} The approach is based on the relation between the
microcanonical entropy, $S(E)=\ln\gee$ (with Boltzmann's constant $k=1$), and the inverse temperature, $\beta(E)=1/T(E)=\partial S/\partial
E$. The main virtue of this method of determining $\gee$ in the present context is that it is an application of the relation between the entropy and the temperature.

For simplicity, we discuss the algorithm for the Ising model for which the values of the energy are discrete. We label the possible energies of the system by the index $j$ with $j=1$ corresponding to the ground state and write the temperature estimate $\tilde T(E_j)$ as $\tilde T_j$. After a trial move to state $j$ we update the entropy $S_j$ as for the Wang-Landau algorithm and also update the temperature estimate at $j \pm 1$. From Eq.~\eqref{gupdate} or
\begin{equation}
\label{eq:updateS}
\tilde S_{j,\,t+1} = \tilde S_{j,\,t}+\ln f,
\end{equation}
we can write the central difference approximation for the inverse temperature $\beta$ as 
\begin{equation}
\frac{\partial \tilde{S}}{\partial E}\Big|_{E=E_{j,t+1}}
= \tilde{\beta}_{j,t+1} \approx
[\tilde S_{j+1,t}-\tilde S_{j-1,t}]/(2\de),
\label{s-temp}
\end{equation}
where $\de$ is the energy difference between state $j$ and states $j\pm 1$. (For the Ising model on a square lattice, $\de = 4J$. In general, we would choose a larger bin size so that several states would correspond to one bin.) The energy $E_j$ in Eq.~\eqref{s-temp} is the value of the energy of the system after a trial move. 
On a visit to $E_j$ we use the updated value of $\tilde{S}_j$ and the unchanged values of $\tilde{S}_{j\pm 2}$ to determine new estimates for the temperature. From Eq.~\eqref{eq:updateS} we have $\tilde{\beta}_{j+1,\,t+1}= [\tilde{S}_{j+2,\,t}-\tilde{S}_{j,t+1}]/(2\Delta E)=\tilde{\beta}_{j+1}-\delta f$, and $\tilde{\beta}_{j-1,\,t+1}= [\tilde{S}_{j,\,t+1}-\tilde{S}_{j-2,\,t}]/(2\Delta E)=\tilde{\beta}_{j-1,\,t}+\delta f$ with $\delta f={\ln f}/{2\de}$. Hence, we obtain
\begin{equation}
\tilde T_{j \pm 1,\,t+1} = 1/\beta_{j\pm 1}= \alpha_{j \pm 1, t}\, \tilde T_{j \pm 1,\,t}, \label{eq:rescaleT}
\end{equation}
where $\alpha_{j \pm 1,t}=1/[1\mp \delta f \tilde T_{j \pm 1,\,t}]$. Note that $\tilde T_{j-1,\,t+1}$ is decreased and $\tilde T_{j+1,\,t-1}$ is increased while $\tilde T_j$ is unchanged. In this way
$\tilde{T}(E)$ will converge to a monotonically increasing function of the energy. It is best to restrict the updates to a finite range of temperatures between $T_{\min}$ and $T_{\max}$.

\begin{figure}[h!]%fig3
%\vspace{-5.25cm}
\includegraphics[scale=0.62]{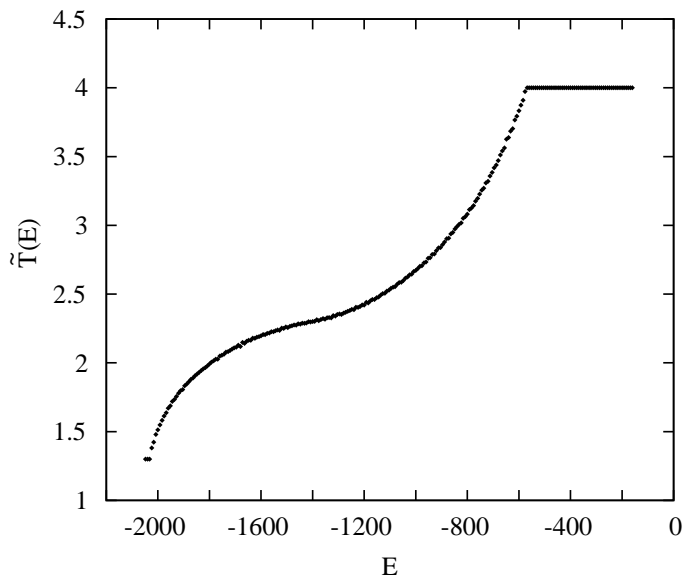}
\vspace{-0.25cm}
\caption{\label{tes} The estimated energy-dependence of the temperature for the Ising model on a $32 \times 32$ square lattice as determined by the method of Ref.~\onlinecite{stmc}. The simulation was done with $T_{\max}= 4.5$ and $T_{\min}=1.3$ using $\de=8$ and converged to $f= 1 + 10^{-12}$ after $4 \times 10^6$ Monte Carlo iterations per spin with the initial modification factor $f_0=1.00005$. The data was generated by Jaegil Kim.}
\end{figure}

The acceptance probability of the trial moves is given by Eq.~\eqref{flat} so we also need to update the entropy as is done for the \WL\ algorithm. The values of $f$ are changed as for the \WL\ algorithm. An example of the converged $\tilde T(E)$ for the Ising model is given in Fig.~\ref{tes}.

Once $\tilde{T}(E)$ converges, the entropy estimate is found by integrating $\tilde{T}(E)$ with respect to $E$. 
In the simplest interpolation method the entropy is given by
\begin{equation}
\label{entropy}
S(E) = \sum_{j=j_{\min}}^j \tilde \beta_j \de.
\end{equation} 
These values of $S(E)$ are then used to obtain $g(E)$, and the thermodynamic properties of the system are determined from Eq.~\eqref{aveE}. Because the continuum entropy $S(E)$ can be obtained by integrating the interpolated $T_j$, the updating of $T(E)$ as in Eq.~\eqref{eq:rescaleT} is especially useful for systems where the energy changes continuously. Even more interesting is the use of Eq.~\eqref{eq:rescaleT} with molecular dynamics simulations at various temperatures.\cite{stmc,stmc2}

\section{Direct measurement of the partition function}

Another \mc\ method combines the Metropolis algorithm with the Wang-Landau method to directly compute the partition function $Z$ at all temperatures of interest.\cite{zhang} The method uses two types of trial moves: a standard Metropolis Monte Carlo move such as a flip of a spin which changes the energy at fixed temperature, and a move to change the temperature at fixed energy. The acceptance rule is given by
\begin{equation} 
A_{i \to j} = \min \Big\{1, \frac{e^{\beta_iE_i}/\tilde Z(\beta_i)}{e^{\beta_jE_j}/\tilde Z(\beta_j)}\Big\},
\label{partacc}
\end{equation}
where $E_i$ is the energy of the current configuration, $\beta_i$ is the current inverse temperature and $\tilde Z(\beta_i)$ is the current estimate of the partition function for inverse temperature $\beta_i$. For an energy move, $\beta_i = \beta_j$, and for a temperature move $E_i = E_j$. At each trial move a decision to choose a temperature instead of an energy move is made with a fixed probability such as 0.1\%. To update the partition function we use the same procedure as for the density of states in the Wang-Landau method:
\begin{equation}
\ln \tilde Z(\beta) \to \ln \tilde Z(\beta) + \ln f.
\label{zcalc}
\end{equation}
This procedure will lead to a stationary state with probability
\begin{equation}
P_i = e^{\beta_i E_i}/Z(\beta_i). 
\label{boltzmann}
\end{equation}

\begin{figure}[h!]%fig4
\includegraphics[scale=0.62]{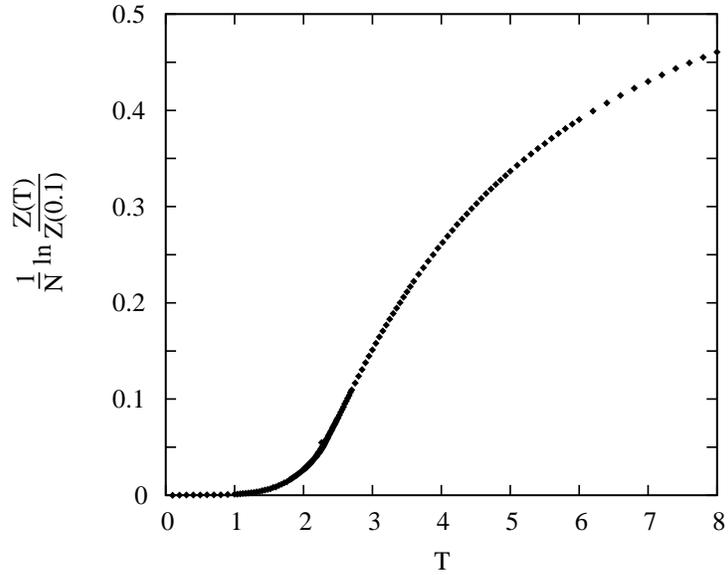}
%\vspace{-4.25cm}
\caption{\label{lnZ} Simulation results for $[\ln Z(T)/Z(0.1)]/N$ for a $32 \times 32$ Ising lattice. $T = 0.1$ is the lowest temperature simulated. The algorithm used and the temperatures simulated are the same as in Ref.~\onlinecite{zhang}. The final value of $\ln f \approx 3.16 \times 10^{-5}$. The number of MC steps at each value of $f$ was $100N/\ln f$.}
\end{figure}

A plot of $\ln{[Z(T)/Z(T=0.1]}/N$ for the two-dimensional Ising model with $N=32 \times 32$ spins is shown in Fig.~\ref{lnZ}. We rarely show plots of partition functions in a thermal physics course. The plot shows that $\ln{Z}$ does not change very much at low temperatures, increases rapidly near the phase transition, and then increases slowly for higher temperatures. 

\section{Summary}
Our focus has been on understanding important concepts in statistical mechanics by considering simulations of concrete models of thermal and statistical systems. Instructors can choose how to use simulations in their courses. For example, students can be asked to write programs with the use of templates.\cite{ejs} Another strategy is to have students run existing programs and modify some of the parameters and explain the results.\cite{stp}

Computers are omnipresent in physics research. Much of this use is for data analysis, symbolic algebra (for example, to calculate Feynman diagrams), and the control of experiments. Statistical physics is an area where the development of new algorithms and simulations has qualitatively changed the kinds of systems and problems that can be considered. These developments make it even more important to think about the ways that computers should change the way we teach thermal and statistical physics.

\section{\label{sec:probs}Suggestions for further study}

\begin{enumerate}
\item Detailed balance.

\begin{enumerate}

\item We showed that the detailed balance condition in Eq.~(\ref{detailedbalance}) ensures that the Boltzmann probability distribution is stationary. Show that the 
detailed balance condition, Eq.~(\ref{generaldetailedbalance}), ensures that the distribution $P_i$ is stationary.\label{prob:desired} 

\item Show that the Metropolis algorithm satisfies detailed balance if $W_{ij} = W_{ji}$. Show that the symmetric acceptance probability, 
\begin{equation} 
A_{i \to j} = \frac{e^{-\beta E_j}}{e^{-\beta E_j} + e^{-\beta E_i}},
\label{symmetric}
\end{equation}
also satisfies detailed balance. \label{prob:metro}

\item Show that the stationary probability distribution for the Wang-Landau algorithm, Eq.~(\ref{flat}), is $P_i = c/g(E_i)$, where $c$ is a constant independent of $E$.\label{prob:flat}

\end{enumerate}

\item Approach to equilibrium.

\begin{enumerate}

\item Write a program that implements the balls and bags example discussed in Sec.~\ref{approach} or use the application/applet available from \url{<stp.clarku.edu>}.\cite{stp}

\item Plot the number of particles on the left side of the container as a function of time. How long does it take for a system of $N=64$ balls to come to equilibrium? Then estimate this time for $N=128$, 256, and 512 balls.

\item Once the system reaches equilibrium there are still fluctuations. Find a relation for the magnitude of the fluctuations as a function of $N$. 

\item Does there exist an initial configuration of the balls that comes to a different equilibrium state than the ones you have simulated so far? Why is the answer to this question important in statistical mechanics? 

\end{enumerate}

\item Wang-Landau algorithm

\begin{enumerate}

\item Calculate by hand the density of states for $N=6$ Ising spins in one dimension with toroidal boundary conditions. The minimum energy is $-6$ and the maximum energy is $+6$. There are $2^6 = 64$ microstates. 

\item Write a program that uses the Wang-Landau algorithm to determine the density of states for the one-dimensional Ising chain and compare your results with your hand calculation.

\end{enumerate}

\item Additional problems related to the topics in this article as well as many other topics in statistical physics can be found at  \url{<stp.clarku.edu>} or EPAPS.\cite{stp} 

\end{enumerate}

\begin{acknowledgments}
We gratefully acknowledge the partial support of National Science Foundation grants DUE-0127363 and DUE-0442481. We thank Jaegil Kim, Jon Machta, and Louis Colonna-Romano for useful discussions and the latter for generating the data in Figs.~\ref{ising123} and \ref{canonical} and generating all the figures. Chris Domenicali wrote the program leading to the data in Fig.~\ref{lnZ}. 
\end{acknowledgments}

\newpage \section*{Figures}

\begin{thebibliography}{5}
\bibitem{interviews}These observations are based on interviews by Jan Tobochnik with students who have taken standard introductory physics and chemistry courses.
\bibitem{noteuler}Although the simple Euler algorithm is frequently used in introductory courses to numerically integrate Newton's equations of motion, it is not adequate for molecular dynamics. The most commonly used algorithm, the Verlet algorithm,\cite{md,gtc} is easy to understand and implement.
\bibitem{densityofstates}In this case as for the Ising model the possible values of $E$ are discrete and hence $\gee$ is the number of states with energy $E$. However, $\gee$ is commonly referred to as the \dos\ independently of whether the energy is a continuous or discrete variable or if the possible energy values are binned.
\end{thebibliography}
\end{document}